\documentclass [12pt, a4paper]{article}
\usepackage {latexsym, amsmath, graphicx}
\usepackage{amssymb}

\begin {document}

\title {Towards Theory of Massive-Parallel Proofs. Cellular Automata Approach}
\author {Andrew Schumann}
\maketitle
\begin{abstract}
In the paper I sketch a theory of massively parallel proofs using
cellular automata presentation of deduction. In this presentation
inference rules play the role of cellular-automatic local
transition functions. In this approach we completely avoid axioms
as necessary notion of deduction theory and therefore we can use
cyclic proofs without additional problems. As a result, a theory
of massive-parallel proofs within unconventional computing is
proposed for the first time.
\end{abstract}
\section{Introduction}
Non-well-founded proofs including cyclic proofs have been actively
studying recently (see \cite{Brotherston1} -- \cite{Brotherston4},
\cite{Santocanale}). Their features consist in that in the
classical theory of deduction, derivation trees, on the one hand,
are finite and, on the other hand, they are without cycles, while
in the non-well-founded approach they can be infinite and, at the
same time, circles occur in them. Non-well-founded proofs have
different applications in computer science. In the paper I am
proposing a more radical approach than other non-well-founded
approaches to deduction by defining massive-parallel proofs and
rejecting axioms in proof theory. This novel approach is
characterized as follows:
\begin{itemize}
    \item Deduction is considered as a transition in cellular
automata, where states of cells are regarded as well-formed
formulas of a logical language.
    \item We build up derivations without using axioms, therefore there is
no sense in distinguishing logic and theory (i.e.\ logical and
nonlogical axioms), derivable and provable formulas, etc.
    \item In deduction we do not obtain derivation trees and instead of the latter
we find out derivation traces, i.e.\ a linear evolution of each
singular premise.
    \item Some derivation traces are circular, i.e.\ some premises
are derivable from themselves.
    \item Some derivation traces are
infinite.
\end{itemize}

\section{Proof-theoretic cellular automata}
For any logical language $\mathcal{L}$ we can construct a
\emph{proof-theoretic cellular automaton }(instead of conventional
deductive systems) simulating massive-parallel proofs.

\newtheorem{definition}{Definition}\begin{definition} A
proof-theoretic cellular automaton is a 4-tuple
$\mathcal{A}=\langle \mathbb{Z}^d$, $S$, $N$, $\delta\rangle$,
where
\begin{itemize}
    \item $d \in \mathbb{N}$ is a number of dimensions and the members of $\mathbb{Z}^d$ are referred as
   cells,
    \item $S$ is a finite or infinite set of elements called the states of
an automaton $\mathcal{A}$, the members of $\mathbb{Z}^d$ take
their values in $S$, the set $S$ is collected from well-formed
formulas of a language $\mathcal{L}$.
    \item $N \subset \mathbb{Z}^d \setminus \{0\}^d$ is a finite
ordered set of $n$ elements, $N$ is said to be a neighborhood,
    \item $\delta \colon S^{n+1} \to S$ that is $\delta$ is the inference rule of a language $\mathcal{L}$, it plays the role of local transition
function of an automaton $\mathcal{A}$.
\end{itemize}
\end{definition}

As we see an automaton is considered on the endless
$d$-dimensional space of integers, i.e. on $\mathbb{Z}^d$.
Discrete time is introduced for $t=0,1,2,\dots$ fixing each step
of inferring.

For any given $z \in \mathbb{Z}^d$, its neighborhood is determined
by $z+N=\{z+\alpha\colon\alpha \in N\}$. There are two often-used
neighborhoods:
\begin{itemize}
\item Von Neumann neighborhood $N_{VN}=\{z \in \mathbb{Z}^d \colon
\sum_{k=1}^d |z_k|=1\}$ \item Moore neighborhood $N_M=\{z \in
\mathbb{Z}^d \colon \max\limits_{k=\overline{1,d}}
|z_k|=1\}=\{-1,0,1\}^d\setminus\{0\}^d$
\end{itemize}

For example, if $d=2$, $N_{VN}=\{(-1,0)$, $(1,0)$, $(0,-1)$,
$(0,1)\}$; $N_M=\{(-1,-1)$, $(-1,0)$, $(-1,1)$, $(0,-1)$, $(0,1)$,
$(1,-1)$, $(1,0)$, $(1,1)\}$.

In the case $d=1$, von Neumann and Moore neighborhoods coincide.
It is easily seen that $|N_{VN}|=2d$, $|N_M|=3^d-1$.

At the moment $t$, the \emph{configuration} of the whole system
(or the \emph{global state}) is given by the mapping $x^t \colon
\mathbb{Z}^d \to S$, and the \emph{evolution} is the sequence
$x^0, x^1, x^2, \dots$ defined as follows:
$x^{t+1}(z)=\delta(x^t(z),x^t(z+\alpha_1),\dots
,x^t(z+\alpha_n))$, where $\langle\alpha_1$, \dots,
$\alpha_n\rangle \in N$. Here $x^0$ is the initial configuration,
and it fully determines the future behavior of the automaton. It
is the set of all premises (not axioms).

We assume that $\delta$ is an inference rule, i.e.\ a mapping from
the set of premises (their number cannot exceed $n = |N|$) to a
conclusion. For any $z \in \mathbb{Z}^d$ the sequence $x^0(z)$,
$x^1(z)$, \dots, $x^t(z)$,\dots is called a \emph{derivation trace
from a state $x^0(z)$}. If there exists $t$ such that
$x^t(z)=x^l(z)$ for all $l>t$, then a derivation trace is
\emph{finite}. It is \emph{circular/cyclic} if there exists $l$
such that $x^t(z)=x^{t+ l}(z)$ for all $t$.

\begin{definition}
In case all derivation traces of a proof-theoretic cellular
automaton $\mathcal{A}$ are circular, this automaton $\mathcal{A}$
is said to be reversible.
\end{definition}

Notice that $x^{t+1}$ depends only upon $x^{t}$, i.e. the previous
configuration. It enables us to build the function $G_
{\mathcal{A}}\colon C_{\mathcal{A}}\to C_{\mathcal{A}}$, where
$C_{\mathcal{A}}$ is the set of all possible configurations of the
cellular automaton $\mathcal{A}$ (it is the set of all mappings
$\mathbb{Z}^d \to S$, because we can take each element of this set
as the initial configuration $x^0$, though not every element can
arise in the evolution of some other configuration).
$G_{\mathcal{A}}$ is called the \emph{global function} of the
automaton.

\newtheorem{example}{Example} \begin{example} [modus ponens] Consider a propositional language
$\mathcal{L}$ that is built in the standard way with the only
binary operation of implication~$\supset$. Let us suppose that
well-formed formulas of that language are used as the set of
states for a proof-theoretic cellular automaton $\mathcal{A}$.
Further, assume that \emph{modus ponens} is a transition rule of
this automaton $\mathcal{A}$ and it is formulated for any
$\varphi$, $\psi \in \mathcal{L}$ as follows:

\[
x^{t+1}(z)=\left\{%
\begin{array}{ll}
    \psi, & \hbox{if $x^t(z) = \varphi \supset \psi$ and $\varphi\in (z+ N)$;} \\
    x^{t}(z), & \hbox{otherwise.} \\
\end{array}%
\right.
\]

The further dynamics will depend on the neighborhood. If we assume
the Moor neighborhood in the 2-dimensional space, this dynamics
will be exemplified by the evolution of cell states in
Fig.~\ref{fig:ad-schu1} -- Fig.~\ref{fig:ad-schu3}.\end{example}

This example shows that first we completely avoid axioms and
secondly we take premisses from the cell states of the
neighborhood according to a transition function. As a result, we
do not come across proof trees in our novel approach to deduction
taking into account that a cell state has just a linear dynamics
(the number of cells and their location do not change). This
allows us evidently to simplify deductive systems.

Now we are trying to consider a cellular-automaton presentation of
two basic deductive approaches: Hilbert's type and sequent ones.

\begin{figure}
\begin{center}
\small
\begin{tabular}{|c|c|c|c|c|}
  \hline
  $(p\supset q)\supset r$ & $p\supset (p\supset q)$ & $p\supset q$ & $(p\supset q)\supset (p\supset q)$ & $(r \supset p)\supset r$ \\
   \hline
  $(p\supset r)\supset (q \supset r)$ & $p\supset q$ & $p$ & $p\supset (p\supset q)$ & $r \supset p$ \\
  \hline
  $p\supset  r$ & $p$ & $p\supset (q \supset (p\supset q))$ & $p$ & $r $ \\
 \hline
  $p\supset (q\supset r)$ & $p\supset p$ & $p\supset q$ & $(p\supset r)\supset (q\supset p)$ & $ p\supset r$ \\
 \hline
  $p\supset q$ & $p\supset (q\supset p)$ & $q$ & $p\supset r$ & $p$ \\

  \hline
\end{tabular}
\end{center}\normalsize
\caption{An initial configuration of a proof-theoretic cellular
automaton $\mathcal{A}$ with the Moor neighborhood in the
2-dimensional space, its states run over formulas set up in a
propositional language $\mathcal{L}$ with the only binary
operation~$\supset$, $t= 0$. Notice that $p,q,r$ are propositional
variables. } \label{fig:ad-schu1}
\end{figure}

\begin{figure}
\begin{center}
\small
\begin{tabular}{|c|c|c|c|c|}
  \hline
  $r$ & $p\supset q$ & $q$ & $p\supset q$ & $r$ \\
   \hline
  $q \supset r$ & $q$ & $p$ & $p\supset q$ & $ p$ \\
  \hline
  $ r$ & $p$ & $q \supset (p\supset q)$ & $p$ & $r $ \\
 \hline
  $q\supset r$ & $ p$ & $q$ & $q\supset p$ & $ r$ \\
 \hline
  $p\supset q$ & $p\supset (q\supset p)$ & $q$ & $ r$ & $p$ \\

  \hline
\end{tabular}
\end{center}\normalsize
\caption{An evolution of $\mathcal{A}$ described in
Fig.~\ref{fig:ad-schu1} at the time step $t=1$.}
\label{fig:ad-schu2}
\end{figure}

\begin{figure}
\begin{center}
\small
\begin{tabular}{|c|c|c|c|c|}
  \hline
  $r$ & $q$ & $q$ & $q$ & $r$ \\
   \hline
  $ r$ & $q$ & $p$ & $ q$ & $ p$ \\
  \hline
  $ r$ & $p$ & $ q$ & $p$ & $r $ \\
 \hline
  $ r$ & $ p$ & $q$ & $ p$ & $ r$ \\
 \hline
  $q$ & $ p$ & $q$ & $ r$ & $p$ \\

  \hline
\end{tabular}
\end{center}\normalsize
\caption{An evolution of $\mathcal{A}$ described in
Fig.~\ref{fig:ad-schu1} at the time step $t=3$. Its configuration
cannot vary further.} \label{fig:ad-schu3}
\end{figure}

\begin{example}[Hilbert's inference rules]
Suppose a propositional language $\mathcal{L}$ contains two basic
propositional operations: negation and disjunction. As usual, the
set of all formulas of $\mathcal{L}$ is regarded as the set of
states of an appropriate proof-theoretic cellular automata. In
that we will use the exclusive disjunction of the following five
inference rules converted from Joseph R.\ Shoenfield's deductive
system:

\[x^{t+1}(z) =\left\{%
\begin{array}{ll}
    \psi\vee\varphi, & \hbox{if $x^t(z) = \varphi$;} \\
    \varphi, & \hbox{if $x^t(z) = \varphi \vee\varphi$;} \\
    (\chi\vee\psi)\vee\varphi, & \hbox{if $x^t(z) = \chi\vee(\psi\vee\varphi)$;} \\
    \chi\vee\psi, & \hbox{if $x^t(z) = \varphi\vee\chi$ and $(\neg\varphi\vee\psi)\,\,\in\,\, (z+N)$;} \\
    \chi\vee\psi, & \hbox{if $x^t (z)= \neg\varphi\vee\psi$ and $(\varphi\vee\chi)\,\,\in\,\, (z+N)$.} \\
\end{array}%
\right.    \]
\end{example}

\begin{example}[sequent inference rules] Let us take a sequent propositional language
$\mathcal{L}$, in which the classical propositional language with
negation, conjunction, disjunction and implication is extended by
adding the sequent relation $\hookrightarrow$. Recall that a
\textit{sequent} is an expression of the form
$\Gamma_1\hookrightarrow \Gamma_2$, where
$\Gamma_1=\{\varphi_1,\dots,\varphi_j\}$,
$\Gamma_2=\{\psi_1,\dots,\psi_i\}$ are finite sets of well-formed
formulas of the standard propositional language, that has the
following interpretation: $\Gamma_1\hookrightarrow \Gamma_2$ is
logically valid iff
$$\bigwedge_j\varphi_j\supset\bigvee_i\psi_i$$ is logically
valid. Let $S$ denote the set of all sequents of $\mathcal{L}$,
furthermore let us assume that this family $S$ is regarded as the
set of states for a proof-theoretic cellular automaton
$\mathcal{A}$. The transition rule of $\mathcal{A}$ is an
exclusive disjunction of the 14 singular rules (6 structural rules
and 8 logical rules):

$$x^{t+1}(z)=\Gamma_1\hookrightarrow \Gamma_2,\left\{%
\begin{array}{ll}
    & \hbox{if $\Gamma_1\hookrightarrow \Gamma_2$ is a result of applying to $x^t(z)$} \\ & \hbox{eather one of structural rules} \\ & \hbox{or the left (right) introduction of negation} \\ & \hbox{or the left introduction of conjunction} \\ & \hbox{or the right introduction of disjunction} \\ & \hbox{or the right introduction of implication.} \\
  \end{array}%
\right.    $$

$$x^{t+1}(z)=\left\{%
\begin{array}{ll}
    \Gamma\hookrightarrow \Gamma',\psi\wedge\chi, & \hbox{if $x^t(z) = \Gamma\hookrightarrow \Gamma',\psi$ and} \\ & \hbox{$(\Gamma\hookrightarrow \Gamma',\chi) \,\, \in \,\, (z+N)$;} \\
    \Gamma,\psi\vee\chi\hookrightarrow \Gamma', & \hbox{if $x^t(z) = \Gamma,\psi\hookrightarrow
\Gamma'$ and} \\ & \hbox{$(\Gamma,\chi\hookrightarrow
\Gamma')\,\, \in \,\, (z+N)$;} \\
    \psi\supset\chi,\Gamma,\Delta\hookrightarrow
\Gamma',\Delta', & \hbox{if $x^t(z) = \Gamma\hookrightarrow
\Gamma',\psi$ and} \\ & \hbox{$(\chi,\Delta\hookrightarrow
\Delta')\,\, \in \,\, (z+N)$.} \\
\end{array}%
\right.    $$

\end{example}

\begin{example}[Brotherston's cyclic proofs] The sequent language used in the previous example we extend by adding
predicates $N$, $E$, $O$ and appropriate inference rules of
Fig.~\ref{fig:ad-schu4} for them. Further, let us extend also the
automaton of Example 3 in the same way by representing inference
rules of Fig.~\ref{fig:ad-schu4} in the cellular-automatic form.

Now we assume that a cell has an initial state $[\Gamma,N(z)
\hookrightarrow \Delta,O(z),E(z)]$ and its neighbor cell an
initial state $[\Gamma, z=0 \hookrightarrow\Delta]$ that is equal
to $[\Gamma, z=0 \hookrightarrow\Delta,O(z),E(z)]$ for any
$t=4,14,24,\dots$ and to $[\Gamma, z=0
\hookrightarrow\Delta,E(z),O(z)]$ for any $t=9,19,29,\dots$. Then
we will have the following infinite cycle:
\\

$[\Gamma, N(z) \hookrightarrow \Delta, O(z),E(z)]
\longrightarrow^{(substitution)} [\Gamma, N(y) \hookrightarrow
\Delta, O(y),E(y)]$ $\longrightarrow [\Gamma, N(y) \hookrightarrow
\Delta, O(y),O(y+1)]$ $\longrightarrow [\Gamma, N(y)
\hookrightarrow \Delta, E(y+1),O(y+1)]$ $\longrightarrow [\Gamma,
z=(y+1),N(y) \hookrightarrow \Delta,
O(z),E(z)]\longrightarrow^{(case \,N)} [\Gamma, N(z)
\hookrightarrow \Delta, E(z),O(z)] \longrightarrow\dots$
\\
\end{example}

Another instance of cyclic proof is given in Example 5. As we see,
the possibility of cyclic derivation traces depends on
configuration of cell states.

Traditional tasks concerning proof theory like completeness and
independence of axioms lose their sense in massive-parallel proof
theory, although it can be readily shown that we can speak about
consistency:

\newtheorem{proposition}{Proposition}\begin{proposition}
Proof theories given in Examples 2 and 3 are consistent, i.e.\ we
cannot deduce a contradiction within them.
\end{proposition}

\begin{figure}
$$\boxed{\begin{array}{c}\begin{array}{cc}\frac{\Gamma \hookrightarrow N(x)}{\Gamma \hookrightarrow N(x+1)},\qquad \frac{\Gamma \hookrightarrow \Delta}{\Gamma \hookrightarrow \Delta, N(0)},\qquad \frac{\Gamma \hookrightarrow E(x)}{\Gamma \hookrightarrow O(x+1)},\qquad \frac{\Gamma \hookrightarrow O(x)}{\Gamma \hookrightarrow E(x+1)},\qquad \frac{\Gamma \hookrightarrow \Delta}{\Gamma \hookrightarrow \Delta, E(0)},\end{array}
\\[1em]\begin{array}{cc}\frac{N(x)\hookrightarrow \Delta}{N(x+1)\hookrightarrow \Delta},\qquad \frac{\Gamma\hookrightarrow \Delta}{\Gamma, N(0)\hookrightarrow \Delta},\qquad \frac{E(x)\hookrightarrow \Delta}{O(x+1)\hookrightarrow \Delta},\qquad \frac{O(x)\hookrightarrow \Delta}{E(x+1)\hookrightarrow \Delta},\qquad \frac{\Gamma\hookrightarrow \Delta}{\Gamma, E(0)\hookrightarrow \Delta},\end{array}
\\[1em]\begin{array}{cc}\frac{\Gamma, t=0\hookrightarrow\Delta\qquad \Gamma, t=x+1, N(x)\hookrightarrow\Delta}{\Gamma, N(t)\hookrightarrow\Delta} \,\,(Case \,\,N), \hbox{where $x \notin FV (\Gamma\cup\Delta\cup\{N(t)\})$,}\end{array}
\\[1em]\begin{array}{cc}\frac{\Gamma \hookrightarrow\Delta}{\Gamma[x] \hookrightarrow\Delta[x]}\,\,(Substitution).\end{array}\end{array}}$$\begin{center}
\end{center}\normalsize
\caption{Inference rules for predicates $N$ (`being a natural
number'), $E$ (`being an even number'), $O$ (`being an odd
number'), see \cite{Brotherston1}.} \label{fig:ad-schu4}
\end{figure}

\section{The proof-theoretic cellular automaton for Belousov-Zhabotinsky reaction}

Massive-parallel computing is observed everywhere in natural
systems. There are different approaches to nature-inspired
computing: reaction-diffusion computing \cite{Adamatzky2} --
\cite{Adamatzky4}, \cite{Schumann2}, chemical computing
\cite{Berry}, biological computing \cite{Ivanitsky},
\cite{Prajer}, etc. In all those computational models parallel
inferring and concurrency are assumed as key notions. In the paper
\cite{Schumann1} a hypothesis was put forward that the paradigm of
parallel and concurrent computation caused by rejecting the
set-theoretic axiom of foundation can be widely applied in modern
physics. In this section we are analyzing simulating
Belousov-Zhabotinsky reaction within the framework of our theory
of massive-parallel proofs.

Let us consider a proof-theoretic cellular automaton with circular
proofs for the Belousov-Zhabo\-tinsky reaction containing feedback
relations. The mechanism of this reaction (namely cerium(III)
$\longleftrightarrow$ cerium(IV) catalyzed reaction) is very
complicated: its recent model contains 80 elementary steps and 26
variable species concentrations. Let us consider a simplification
of Belousov-Zhabotinsky reaction assuming that the set of states
consists just of the following reactants: $Ce^{3+}$, $HBrO_2$,
$BrO_3^-$, $H^+$, $Ce^{4+}$, $H_2O$, $BrCH$ $(COOH)_2$, $Br^-$,
$HCOOH$, $CO_2$, $HOBr$, $Br_2$, $CH_2(COOH)_2$ which interact
according to inference rules (reactions) \eqref{SchumannEq1} --
\eqref{SchumannEq7}. In this reaction we observe sudden
oscillations in color from yellow to colorless, allowing the
oscillations to be observed visually. In spatially nonhomogeneous
systems (such as a simple petri dish), the oscillations propagate
as spiral wave fronts. The oscillations last about one minute and
are repeated over a long period of time. The color changes are
caused by alternating oxidation-reductions in which cerium changes
its oxidation state from cerium(III) to cerium(IV) and vice versa:
$Ce^{3+} \longrightarrow Ce^{4+} \longrightarrow Ce^{3+}
\longrightarrow\dots$.

When $Br^-$ has been significantly lowered, the reaction pictured
by inference rule \eqref{SchumannEq1} causes an exponential
increase in bromous acid ($HBrO_2$) and the oxidized form of the
metal ion catalyst and indicator, cerium(IV). Bromous acid is
subsequently converted to bromate ($BrO_3^-$) and $HOBr$ (the step
\eqref{SchumannEq3}). Meanwhile, the step \eqref{SchumannEq2}
reduces the cerium(IV) to cerium(III) and simultaneously increase
bromide ($Br^-$) concentration. Once the bromide concentration is
high enough, it reacts with bromate ($BrO_3^-$) and $HOBr$ in
\eqref{SchumannEq4} and \eqref{SchumannEq6} to form $Br_2$,
further $Br_2$ reacts with $CH_2(COOH)_2$ to form $BrCH(COOH)_2$
and the process begins again. Thus, parallel processes in
\eqref{SchumannEq1} -- \eqref{SchumannEq7} have several cycles
which are performed synchronously.

The proof-theoretic simulation of Belousov-Zhabotinsky reaction
can be defined as follows:

\begin{definition} Consider a propositional language
$\mathcal{L}$ with the only binary operation $\oplus$, it is built
in the standard way over the set of variables $S = \{Ce^{3+}$,
$HBrO_2$, $BrO_3^-$, $H^+$, $Ce^{4+}$, $H_2O$, $BrCH$ $(COOH)_2$,
$Br^-$, $HCOOH$, $CO_2$, $HOBr$, $Br_2$, $CH_2(COOH)_2\}$. Let $S$
be the set of states of proof-theoretic cellular automaton
$\mathcal{A}$. The inference rule of the automaton is presented by
the conjunction of singular inference rules \eqref{SchumannEq1} --
\eqref{SchumannEq7}:

$$\eqref{SchumannEq1} \wedge\eqref{SchumannEq2} \wedge\eqref{SchumannEq3} \wedge\eqref{SchumannEq4} \wedge\eqref{SchumannEq5} \wedge\eqref{SchumannEq6} \wedge \eqref{SchumannEq7}.$$

The operation $\oplus$ has the following meaning: $ A\oplus B$
defines a probability distribution of events $A$ and $B$ in
neighbor cells participated in a reaction caused the appearance of
$ A\oplus B$. Then $\mathcal{A}$ simulates the
Belousov-Zhabo\-tinsky reaction.\end{definition}
\begin{definition} Let $p, s_i, s_{i+1}\in\{Ce^{3+}$,
$HBrO_2$, $BrO_3^-$, $H^+$, $Ce^{4+}$, $H_2O$, $BrCH$ $(COOH)_2$,
$Br^-$, $HCOOH$, $CO_2$, $HOBr$, $Br_2$, $CH_2(COOH)_2\}$. A state
$p$ is called a premise for deducing $s_{i+1}$ from $s_i$ by the
inference rule $\eqref{SchumannEq1} \wedge\eqref{SchumannEq2}
\wedge\dots \wedge \eqref{SchumannEq7}$ iff
\begin{itemize}
    \item $p$ is $s_i$ or
    \item in a neighbor cell we find out an expression of the form $p \oplus
A, B\oplus C$, where $A, B, C$ are propositional metavariables,
i.e.\ they run over either the empty set or the set of states
closed under the operation $\oplus$. Thus, we assume that each
premise should occur in a separate cell. This means that if we
find out an expression $p_i \oplus p_j, B\oplus C$ or $p_i \oplus
A, p_j\oplus C$ in a neighbor cell and both $p_i$ and $p_j$ are
needed for deducing, whereas $p_i$, $p_j$ do not occur in other
neighbor cells, then $p_i$, $p_j$ could not be considered as
premises.
\end{itemize}
\end{definition}

\begin{equation}x^{t+1}(z) =\left\{%
\begin{array}{ll}
    (1)\,\,Ce^{4+} \oplus HBrO_2 \oplus H_2O, \,\,\, \hbox{if $x^t(z) \in\{
    Ce^{3+}\}$ and}\\ \,\,\,\, \hbox{premises $HBrO_2, BrO_3^-,
H^+\in \,(z+ N)$;} \\
    (2)\,\, x^t(z),  \,\,\,\hbox{otherwise.} \\
\end{array}%
\right.\label{SchumannEq1}\end{equation}

\begin{equation}x^{t+1}(z)=\left\{%
\begin{array}{ll}
   (1)\,\, Br^- \oplus Ce^{3+}\oplus HCOOH\oplus CO_2\oplus H^+, \,\,\, \hbox{if $x^t(z)\in
    \{Ce^{4+}\}$} \\  \hbox{and premises $ BrCH(COOH)_2$, $ H_2O\in\, (z+ N)$;} \\
   (2)\,\,  x^t (z), \,\,\, \hbox{otherwise.} \\
\end{array}%
\right. \label{SchumannEq2}\end{equation}

\begin{equation}x^{t+1}(z) =\left\{%
\begin{array}{ll}
   (1)\,\, HOBr \oplus BrO_3^- \oplus H^+, & \hbox{if $x^t(z)\in \{HBrO_2\}$;} \\
   (2)\,\,  x^t(z), & \hbox{otherwise.} \\
\end{array}%
\right.\label{SchumannEq3}
\end{equation}

\begin{equation}x^{t+1}(z)=\left\{%
\begin{array}{ll}
   (1)\,\, HOBr \oplus HBrO_2, & \hbox{if $x^t (z)\in \{BrO^-_3\}$ and} \\ & \hbox{premises $Br^- , H^+\in \,(z+N)$;} \\
   (2)\,\,  x^t(z), & \hbox{otherwise.} \\
\end{array}%
\right.\label{SchumannEq4}
\end{equation}

\begin{equation}x^{t+1}(z)=\left\{%
\begin{array}{ll}
   (1)\,\, HOBr, & \hbox{if $x^t(z) \in \{Br^-\}$ and} \\ & \hbox{premises $HBrO_2, H^+\in\,(z+ N)$;} \\
    (2)\,\, x^t(z), & \hbox{otherwise.} \\
\end{array}%
\right.\label{SchumannEq5}
\end{equation}

\begin{equation}x^{t+1}(z)=\left\{%
\begin{array}{ll}
    (1)\,\,Br_2 \oplus H_2O, & \hbox{if $x^t(z) \in \{HOBr\}$ and} \\ & \hbox{premises $Br^-, H^+\in \,(z+N)$;} \\
    (2)\,\, x^t(z), & \hbox{otherwise.} \\
\end{array}%
\right.\label{SchumannEq6}
\end{equation}

\begin{equation}x^{t+1}(z)=\left\{%
\begin{array}{ll}
   (1)\,\, Br^- \oplus H^+ \oplus
BrCH(COOH)_2, \,\,\, \hbox{if $x^t(z) \in \{Br_2\}$}  \\ \hbox{and premises $CH_2(COOH)_2\in \, (z+N)$;} \\
    (2)\,\, x^t(z), \,\,\, \hbox{otherwise.} \\
\end{array}%
\right. \label{SchumannEq7}\end{equation}

\begin{figure}
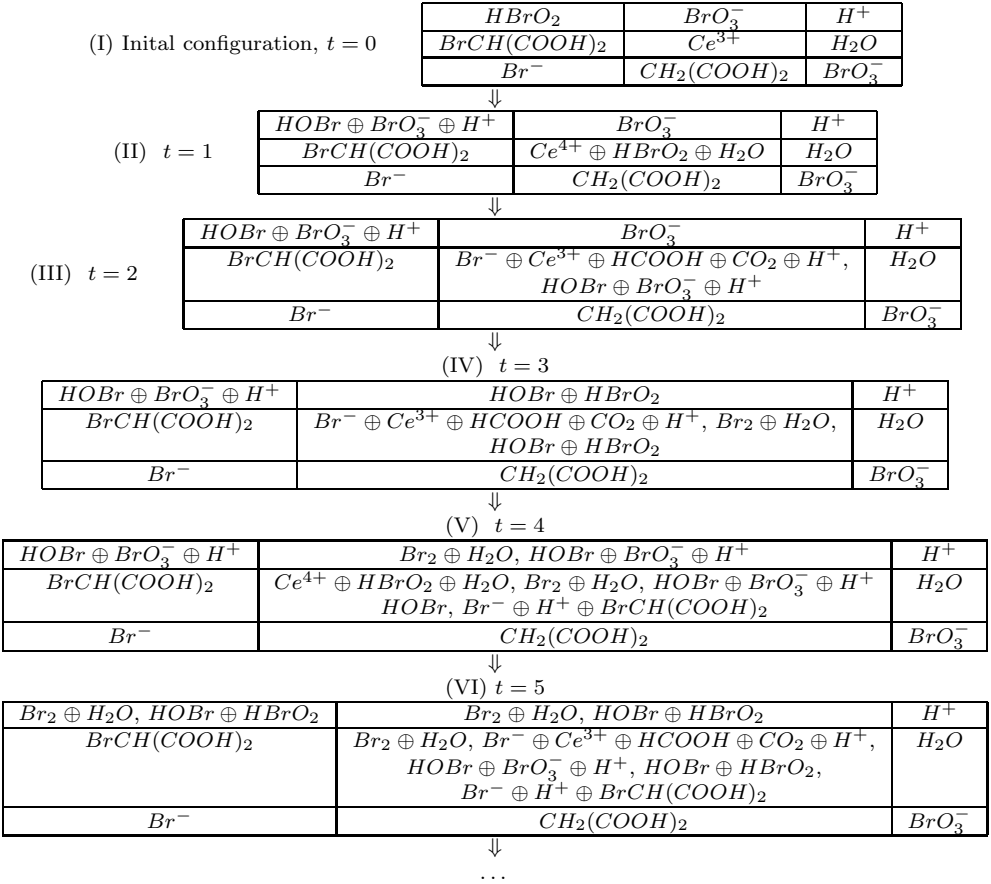

\begin{center}
\scriptsize (I)\,\,Inital configuration,
$t=0$\qquad\begin{tabular}{|c|c|c|c|c|}
  \hline
  $HBrO_2$ & $BrO_3^-$ & $H^+$  \\
   \hline
   $BrCH(COOH)_2$ & $Ce^{3+}$ & $H_2O$ \\
  \hline
 $Br^-$ & $CH_2(COOH)_2$ & $BrO_3^-$\\
 \hline
\end{tabular}

$\Downarrow$\\

(II)\,\, $t=1$\qquad\begin{tabular}{|c|c|c|c|c|}
  \hline
  $HOBr \oplus BrO_3^- \oplus H^+$ & $BrO_3^-$ & $H^+$  \\
   \hline
  $BrCH(COOH)_2$ & $Ce^{4+} \oplus HBrO_2 \oplus
H_2O$ & $H_2O$ \\
  \hline
  $Br^-$ & $CH_2(COOH)_2$ & $BrO_3^-$\\
 \hline
\end{tabular}

$\Downarrow$\\

(III)\,\, $t=2$\qquad\begin{tabular}{|c|c|c|c|c|}
  \hline
  $HOBr \oplus BrO_3^- \oplus H^+$ & $BrO_3^-$ & $H^+$  \\
   \hline
  $BrCH(COOH)_2$ & $Br^-\oplus Ce^{3+}\oplus HCOOH\oplus CO_2\oplus
H^+$, & $H_2O$ \\
&$HOBr \oplus BrO_3^- \oplus H^+$ &\\
  \hline
   $Br^-$ & $CH_2(COOH)_2$ & $BrO_3^-$\\
 \hline
\end{tabular}

$\Downarrow$\\

(IV)\,\, $t=3$\qquad\begin{tabular}{|c|c|c|c|c|}
  \hline
  $ HOBr \oplus BrO_3^- \oplus H^+$ & $HOBr \oplus HBrO_2$ & $H^+$  \\
   \hline
  $BrCH(COOH)_2$ & $Br^-\oplus Ce^{3+}\oplus HCOOH\oplus CO_2\oplus
H^+$, $Br_2 \oplus H_2O$, & $H_2O$ \\
& $HOBr \oplus HBrO_2$ &
\\
  \hline
   $Br^-$ & $CH_2(COOH)_2$ & $BrO_3^-$\\
 \hline
\end{tabular}

$\Downarrow$\\

(V)\,\, $t=4$\qquad\begin{tabular}{|c|c|c|c|c|}
  \hline
  $HOBr \oplus BrO_3^- \oplus H^+$ & $Br_2 \oplus H_2O$, $HOBr \oplus BrO^-_3 \oplus H^+$ & $H^+$  \\
   \hline
  $BrCH(COOH)_2$ & $Ce^{4+}\oplus HBrO_2\oplus H_2O$, $Br_2 \oplus H_2O$, $ HOBr \oplus BrO_3^- \oplus H^+$ & $H_2O$ \\
&$HOBr$, $Br^- \oplus H^+ \oplus BrCH(COOH)_2$&\\
  \hline
   $Br^-$ & $CH_2(COOH)_2$ & $BrO_3^-$\\
 \hline
\end{tabular}

$\Downarrow$\\

(VI) $t=5$\qquad\begin{tabular}{|c|c|c|c|c|}
  \hline
  $Br_2 \oplus H_2O$, $HOBr \oplus HBrO_2$ & $Br_2 \oplus H_2O$, $HOBr \oplus HBrO_2$ & $H^+$  \\
   \hline
  $BrCH(COOH)_2$ & $Br_2 \oplus H_2O$, $Br^-\oplus Ce^{3+}\oplus HCOOH\oplus CO_2\oplus
H^+$,  & $H_2O$ \\
&$HOBr\oplus BrO_3^-\oplus H^+$, $HOBr \oplus HBrO_2$,&\\
&$Br^- \oplus H^+ \oplus BrCH(COOH)_2$&\\  \hline
   $Br^-$ & $CH_2(COOH)_2$ & $BrO_3^-$\\
 \hline
\end{tabular}

$\Downarrow$\\
\dots

\normalsize
\end{center}

\caption{The evolution of a reversible proof-theoretic cellular
automaton $\mathcal{A}$ with the Moor neighborhood in the
2-dimensional space for the Belousov-Zhabotinsky reaction. This
automaton simulates the circular feedback $Ce^{3+} \longrightarrow
Ce^{4+} \longrightarrow Ce^{3+} \longrightarrow\dots$ (more
precisely temporal oscillations in a well-stirred solution):
$Ce^{3+}$ is colorless and $Ce^{4+}$ is yellow. The initial
configuration of $\mathcal{A}$-cells described in (I) occurs in
the same form at the further steps and the cycle repeats several
times. For entailing (I) $\longrightarrow$ (II) we have just used
inference rule \eqref{SchumannEq1} (row 2, column 2) and inference
rule \eqref{SchumannEq3} (row 1, column 1), for entailing (II)
$\longrightarrow$ (III) inference rules \eqref{SchumannEq2},
\eqref{SchumannEq3} and \eqref{SchumannEq8} (row 2, column 2), for
entailing (III) $\longrightarrow$ (IV) inference rule
\eqref{SchumannEq4} (row 1, column 2) and inference rules
\eqref{SchumannEq4}, \eqref{SchumannEq6}, \eqref{SchumannEq8} and
\eqref{SchumannEq9} (row 2, column 2), for entailing (IV)
$\longrightarrow$ (V) inference rules \eqref{SchumannEq3} and
\eqref{SchumannEq6} (row 1, column 2) inference rules
\eqref{SchumannEq1}, \eqref{SchumannEq3}, \eqref{SchumannEq5},
\eqref{SchumannEq6}, \eqref{SchumannEq7}, \eqref{SchumannEq8} and
\eqref{SchumannEq9} (row 2, column 2), for entailing (V)
$\longrightarrow$ (VI) inference rules \eqref{SchumannEq4},
\eqref{SchumannEq6} (row 1, column 1), inference rules
\eqref{SchumannEq4}, \eqref{SchumannEq6}, \eqref{SchumannEq10}
(row 1, column 2), inference rules \eqref{SchumannEq2},
\eqref{SchumannEq3}, \eqref{SchumannEq4}, \eqref{SchumannEq6},
\eqref{SchumannEq7}, \eqref{SchumannEq9}, \eqref{SchumannEq10},
\eqref{SchumannEq11} (row 2, column 2).} \label{fig:ad-schu5}
\end{figure}

\begin{example}[Belousov-Zhabotinsky's cyclic proofs] We can simplify the automaton defined above
assuming that $\oplus$ is a metatheoretic operation with the
following operational semantics:
\[\frac{A \oplus B}{A}\qquad \frac{A \oplus B}{B},\]
where $A$ and $B$ are metavariables defined on $S$. The informal
meaning of that operation is that we can ignore one of both
variables coupled by $\oplus$. In the cellular automaton
$\mathcal{A}$ this metaoperation will be used as follows: \begin{equation}x^{t+1}(z) = \left\{%
\begin{array}{ll}
    X, Y, & \hbox{if $x^t(z)= A \oplus B$ and according to rules \eqref{SchumannEq1} -- \eqref{SchumannEq7},}  \\ &\hbox{$X$ changes from $A$ and $Y$ changes from $B$;}\\
    X, & \hbox{if $x^t(z)= A \oplus B$ and according to rules \eqref{SchumannEq1} -- \eqref{SchumannEq7},}  \\ &\hbox{$X$ changes from $A$ and $B$ does not change;} \\
    Y, & \hbox{if $x^t(z)= A \oplus B$ and according to rules \eqref{SchumannEq1} -- \eqref{SchumannEq7},}  \\ &\hbox{$Y$ changes from $B$ and $A$ does not change;} \\
A \oplus B, & \hbox{if $x^t(z)= A \oplus B$ and rules \eqref{SchumannEq1} -- \eqref{SchumannEq7}}  \\ &\hbox{cannot be applied to $A$ or $B$.}\\

\end{array}%
\right.   \label{SchumannEq8} \end{equation} Let us suppose now
that $X,Y$ run over the set of states closed under the operation
$\oplus$.
\begin{equation}x^{t+1}(z) = \left\{%
\begin{array}{ll}
    X, Y, & \hbox{if (i) $x^t(z)= X, Y$ and (ii) both $X$ and $Y$}  \\ &\hbox{are simultaneously usable (not usable)}  \\ &\hbox{as premises in at least two different rules}  \\ &\hbox{of \eqref{SchumannEq1} --
    \eqref{SchumannEq7} \hbox{(see definition 4)};}\\
    X, & \hbox{if (i) $x^t(z)= X, Y$ and (ii) only $X$ is usable}  \\ &\hbox{as a premise in at least one rule of \eqref{SchumannEq1} --
    \eqref{SchumannEq7};} \\
    Y, & \hbox{if (i) $x^t(z)= X, Y$ and (ii) only $Y$ is usable}  \\ &\hbox{as a premise in at least one rule of \eqref{SchumannEq1} --
    \eqref{SchumannEq7}.} \\
\end{array}%
\right.  \label{SchumannEq9}  \end{equation}

\begin{equation} \hbox{idempotency: $ A ::= A ,A$}.\label{SchumannEq10}\end{equation}
\begin{equation} \hbox{commutativity: $ A, B ::= B ,A$}.\label{SchumannEq11}\end{equation}

Hence, we cannot ignore one of both variables coupled by $\oplus$
and should accept both them if in the neighborhood there are
reactants that catenate both variables and change them. This rule
is the simplest interpretation of $ A\oplus B$ in definition 3. We
have three cases: (i) both variables are catenated with reactants
from the neighborhood, in this case we mean that the probability
distribution of events $A$ and $B$ is the same and equal to 0.5
and, as a result, we cannot choose one of them and accept both;
(ii) only $A$ is catenated with reactants from the neighborhood,
then the probability distribution of event $A$ is equal to 1.0 and
that of $B$ to 0.0; (iii) only $B$ is catenated with reactants
from the neighborhood, then the probability distribution of event
$B$ is equal to 1.0 and that of $A$ to 0.0. Thus, $ A\oplus B$ is
a function that associates either exactly one value with its
arguments (i.e.\ either $A$ or $B$) or simultaneously both values
(i.e.\ $A$ and $B$).

This simplified version of the automaton $\mathcal{A}$ is
exemplified in Fig.~\ref{fig:ad-schu5}.
\end{example}

Evidently, reducing the complicated dynamics of
Belousov-Zhabotinsky reaction to conventional logical proofs is a
task that cannot be solved in easy way differently from simulating
within massive-parallel proofs.

\section{Conclusion}

In this paper we have considered a possibility of consistent proof
theory in that there are no axioms or axiom schemata.

\begin {thebibliography} {text}
\bibitem{Adamatzky2} Adamatzky A. \emph{Computing in
Nonlinear Media and Automata Collectives}. Institute of Physics
Publishing, 2001. \bibitem{Adamatzky3} Adamatzky A., De Lacy
Costello B., Asai T. \emph{Reaction-Diffusion Computers},
Elsevier, 2005.
\bibitem{Adamatzky4} Adamatzky A., A. Wuensche, and B. De Lacy
Costello, Glider-based computation in reaction-diffusion hexagonal
cellular automata, \emph{Chaos, Solitons $\&$ Fractals} 27, 2006,
287--295.
\bibitem{Berry} Berry G., Boudol G. The chemical abstract machine, \emph{Teor. Comput.
Sci}., 96, 1992, 217--248.
\bibitem{Brotherston1} Brotherston J. Cyclic proofs for first-order logic with inductive definitions [in:] B. Beckert, editor, \emph{TABLEAUX 2005}, volume 3702 of LNAI, Springer-Verlag, 2005, 78--92.
\bibitem{Brotherston3} Brotherston J. \emph{Sequent Calculus Proof Systems for Inductive Definitions}. PhD thesis, University of Edinburgh, November 2006.
\bibitem{Brotherston4} Brotherston J. Simpson, A., Complete sequent calculi for induction and infinite descent. \emph{LICS-22}, IEEE Computer Society, July 2007, 51--60.
\bibitem{Ivanitsky} Ivanitsky G. R., Kunisky A. S., Tzyganov M.
A. Study of `target patterns' in a phage-bacterium system,
\emph{Self-organization: Autowaves and Structures Far From
Equilibrium}. Ed. V.I. Krinsky. Heidelberg-Springer, 1984,
214--217.
\bibitem{Schumann1}
Khrennikov A., Schumann A. Physics Beyond The Set-Theoretic Axiom
of Foundation, [in:] \emph{AIP Conf. Proc}. -- March 10, 2009 --
Volume 1101. 374--380.
\bibitem{Prajer} Prajer M., Fleury A., Laurent M. Dynamics
of calcium regulation in Paramecium and possible morphogenetic
implication, \emph{Journal Cell Sci}., 110, 1997, 529--535.
\bibitem{Santocanale} Santocanale L., A calculus of circular proofs and its categorical semantics, [in:] M. Nielsen and U. Engberg, editors, \emph{Proc. of FoSSaCS 2002}, Grenoble, Apr. 2002, Springer-Verlag LNCS 2303, 357--371.
\bibitem{Schumann2} Schumann A., Adamatzky A.
Towards Semantical Model of Reaction-Diffusion Computing,
\emph{Kybernetes}, 38 (9), 2009, pp. 1518 - 1531.
\bibitem{Schumann2a} Schumann A., Adamatzky A. Physarum Spatial
Logic, \emph{New Mathematics and Natural Computation}, 2010 (to
appear).
\bibitem{Schumann3} Schumann A.
Non-well-founded probabilities on streams, [in:] D. Dubois et al.,
editors, \emph{Soft Methods for Handling Variability and
Imprecision}, Advances in Soft Computing 48, 2008, 59--65.
\end
{thebibliography}

Andrew Schumann

Department of Philosophy and Science Methodology,

Belarusian State University, Minsk, Belarus

e-mail: Andrew.Schumann@gmail.com

\end{document}